\documentclass[12pt]{article}

\usepackage{epsfig}
\newcommand{\cO}{\mathcal{O}}

\newcommand{\AmS}{{\protect\the\textfont2
  A\kern-.1667em\lower.5ex\hbox{M}\kern-.125emS}}

\begin{document}
\noindent
 

DESY 02 - 100

Edinburgh 2002/05

LU-ITP 2002/014

\begin{center}

{\huge \bf 
A lattice study of the spin structure of the $\Lambda$ hyperon}

\vskip 1 cm
\normalsize
M. G{\"o}ckeler$^{\rm a,b}$, R. Horsley$^{\rm c}$, D. Pleiter$^{\rm d}$,
P.E.L. Rakow$^{\rm b}$,
S. Schaefer$^{\rm b}$,\\
A. Sch{\"a}fer$^{\rm b}$,
and G.~Schierholz$^{\rm d,e}$\\ 
- QCDSF Collaboration -\\
\vskip 0.5 cm

$^{\rm a}$ Institut f{\"u}r Theoretische Physik,
     Universit{\"a}t Leipzig,\\ 
D-04109 Leipzig, Germany,\\
$^{\rm b}$ Institut f{\"u}r Theoretische Physik,
     Universit{\"a}t Regensburg, \\
D-93040 Regensburg, Germany,\\
$^{\rm c}$ Department of Physics and Astronomy, University of Edinburgh,
Edinburgh EH9 3JZ, Scotland, UK\\
$^{\rm d}$ John von Neumann--Institut f{\"u}r Computing NIC,\\
        D-15735 Zeuthen, Germany\\
$^{\rm e}$ Deutsches Elektronen-Synchrotron DESY,
      D-22603 Hamburg, Germany \\
 
\vskip 0.5 cm
Abstract:\\
\end{center}

The internal spin structure of the $\Lambda$ is of special importance for 
the understanding of the spin structure of hadrons in general.
The comparison between 
the nucleon and $\Lambda$ allows for a test of 
the relevant flavour-symmetry breaking effects. 
Using nonperturbatively $O(a)$ improved Wilson fermions in the 
quenched approximation we have calculated the first moments of the 
unpolarised,
longitudinally polarised and transversity quark distribution functions 
 in the $\Lambda$. The results indicate that flavour symmetry breaking has 
little effect on the internal spin structure, in accordance with model based 
expectations.


\newpage 
\section{INTRODUCTION}
The detailed investigation of the nucleon spin structure 
during the last few years has fuelled  a large number of efforts to 
understand the results with quark models of various kinds. 
It turned out during the course of these investigations that 
SU(3) flavour breaking is of central importance for any such 
effort. A small selection of relevant articles can be found in 
\cite{r1}. Presently the different predictions  
cover a rather broad range, and while some authors claim great accuracy,
others conclude that the existing experimental data 
on hyperon decays is insufficient for any stringent prediction
\cite{goeke}. 
The $\Lambda$ spin structure, in turn, is  
especially sensitive to flavour SU(3) breaking
\cite{jaffe,r2,ashery}.
While in the naive SU(6) quark model the spin of the $\Lambda$ 
is carried exclusively by the $s$ quarks, the SU(3) rotated 
results for the nucleon spin structure suggest that the $s$ and $\bar{s}$
quarks
carry only $\approx 60 \%$ of the $\Lambda$ spin 
while $u$, $\bar{u}$, $d$ and $\bar{d}$ quarks
quarks contribute $\approx - 40 \%$. 
Any specific assumptions about SU(3) symmetry breaking 
will strongly affect these conclusions.
Obviously  additional input is needed, some of which
we want to provide with the  lattice results presented in this paper.\\
The spin structure of the $\Lambda$
is also experimentally of special interest, because its polarisation can 
easily be 
measured via the self-analysing weak decay $\Lambda \to p \pi^-$.
Indeed, the $\Lambda$ polarisation has been determined
at the $Z^0$ pole in $e^+ e^-$ annihilation~\cite{lep} where it is
mainly due to the $s$ quarks, which when produced via $Z^0$ decays have 
an average polarisation of $-0.91$. Furthermore, $\Lambda$ polarisation
has been studied in deep-inelastic scattering of polarised positrons on 
unpolarised protons in the current fragmentation region, i.e.\ selecting
$\Lambda$s which most likely originate from the struck quark~\cite{hermes}.
Needless to say, however, that the interpretation of the experiments 
is complicated by fragmentation effects, $\Lambda$s from the decay of 
heavier hyperons, etc. 
$\Lambda$ and $\overline \Lambda$ polarisation was also analysed for
charged-current neutrino nucleon reactions \cite{naumov}.
A completely unsettled puzzle is the
very strong polarisation of $\Lambda$s produced in unpolarised 
$p+N$ reactions \cite{pN}. 
The knowledge of the internal spin structure 
of the $\Lambda$ is certainly a necessary ingredient for any explanation.\\
Thus, there is ample motivation to perform lattice calculations 
which provide information on the internal $\Lambda$ structure.
For a spin 1/2 baryon
the fraction $\Delta q$ of the spin 
carried by the quarks (and antiquarks) of flavour $q$ is given
in terms of the forward matrix element of the axial vector current:
\begin{equation}
\langle p,s | \bar{q} \gamma_\mu \gamma_5 q | p,s \rangle
 = 2 s_\mu \Delta q \,,
\end{equation}
where $s^2 = - M^2$ ($M=$ mass of the baryon) 
and the states are normalised according to 
$\langle p,s |p',s' \rangle = 2 E_p (2 \pi)^3 
 \delta(\vec{p} - \vec{p}^{\,\prime}) \delta_{s s'}$.
Simultaneously we have calculated the transversity matrix element
(the tensor charges $\delta q$): 
\begin{equation}
\langle p,s | \bar{q} ~{\rm i} \sigma_{\mu\nu} \gamma_5~ q | p,s \rangle
 = {2 \over M} (s_{\mu} p_{\nu} - s_{\nu} p_{\mu}) \delta q  \,.
\end{equation}
The physical origin of the difference between $\Delta q$ and $\delta q$
is that in the first case the baryon is boosted along its spin direction
and in the second case perpendicular to it.
Thus this difference has a very specific sensitivity to
details of the internal baryon structure, being e.g. zero in
the non-relativistic limit.
Operators with additional derivatives can be used to extract also 
some higher moments of the corresponding distribution functions.\\
Our aim is to study how strongly the mass difference between the 
strange quark and the light quarks affects the flavour SU(3) symmetry 
between the (polarised) distribution functions in the nucleons and 
the $\Lambda$. Perfect SU(3) symmetry predicts the equations
\begin{eqnarray}
&\Delta s_{\Lambda} &= (2\Delta u_p-\Delta d_p+2\Delta s_p)/3 ~~~,
\nonumber \\
\Delta d_{\Lambda} = &\Delta u_{\Lambda} &=
(\Delta u_p+4\Delta d_p+\Delta s_p)/6
\end{eqnarray}
and the same relations for the tensor charges.\\
A preliminary account of some of our results has already been given in 
Ref. \cite{meinulf8}.

\section{THE SIMULATION}

We have performed quenched simulations with the Wilson gauge action 
at $\beta = 6.0$ 
using nonperturbatively $O(a)$ improved fermions (clover fermions) 
with $c_{\mathrm {SW}} = 1.769$. The lattice size was $16^3 \times 32$.
We worked with nine combinations of hopping parameters:
$\kappa_u = \kappa_d$, $\kappa_s$ $\in \{0.1324, 0.1333, 0.1342 \} $
corresponding to bare quark masses of 
$\approx 166$, $112$, $58$ MeV, respectively. So we can
{\em extra}\hspace{0.5pt}polate 
to the chiral limit in $\kappa_u = \kappa_d$ and 
{\em inter}\hspace{0.8pt}polate 
in $\kappa_s$ to the physical value $\kappa_s^*$.
For the critical hopping parameter $\kappa_c$ we take the value 0.135201
determined from the PCAC quark mass~\cite{dirk}, 
$\kappa_s^* = 0.1341$ was fixed by requiring the pseudoscalar mass 
$m_{\mathrm {PS}}$ for 
$\kappa_u = \kappa_d = \kappa_c$ and $\kappa_s = \kappa_s^*$ to
be equal to the $K^+$ mass of $494$ MeV (with the scale set by the 
force parameter $r_0 = 0.5$ fm). In the extra- and interpolation 
we assumed a linear dependence of $m_{\mathrm {PS}}^2$ on the quark mass.\\
As an interpolating field for the $\Lambda$ we used (employing Euclidean
notation from now on)
\begin{equation}
 \Lambda_{\alpha}(t) = 
\sum_{x, \, x_4 = t} \epsilon_{i j k} s_{i\alpha} (x) 
\left( u_j^T (x) C \gamma_5  d_k (x) \right)
\end{equation}
where $C$ is the charge conjugation matrix;
$i$, $j$, and  $k$ are colour indices and 
$\alpha$ is a Dirac index.
The quark fields were (Jacobi) smeared to improve the overlap with the 
$\Lambda$ state.\\
As usual, the bare matrix elements are determined from ratios of three-point
functions and two-point functions 
\begin{equation}
R_{\mathcal {O}} = 
\frac{\tilde\Gamma_{\beta \alpha} \langle 
\Lambda_\alpha(t) \mathcal{O}(\tau) \bar \Lambda_\beta(0) \rangle}
         {\Gamma_{\beta \alpha}\langle \Lambda_\alpha(t)  \bar 
\Lambda_\beta(0) \rangle} ~~~, ~~~ 0<\tau<t
\end{equation}
with appropriate spin projection matrices $\Gamma$ and 
$\tilde\Gamma$. The choice for $\Gamma$ and $\tilde\Gamma$
as well as $t$ and the smearing parameters can be found in  \cite{meinulf5}.
The quality of the data can be judged by how
well pronounced a plateau is obtained for this ratio as function 
of $\tau$.\\
On the lattice the choice of the operators 
$\mathcal{O}$ is a non-trivial task, because
the discretisation reduces the symmetry group of (Euclidean) 
space-time from $O(4)$ to 
the hypercubic group $H(4) \subset  O(4)$. Therefore, 
one has to find combinations of operators which avoid the problem
of operator mixing \cite{meinulf1}. 
After renormalisation, the 
matrix elements of these 
operators are expressed  in terms 
of the reduced matrix elements
$v_2$, $a_0$, $a_1$ and $t_0$ 
which correspond to moments of the parton distribution functions.
We work with the operators
\begin{equation}
\cO_{v_2} = {1\over 2} ~
 \bar q \left( \gamma_4 \stackrel{\leftrightarrow}D_4
- {1\over 3} \left( \gamma_1 \stackrel{\leftrightarrow}D_1
+\gamma_2 \stackrel{\leftrightarrow}D_2
+\gamma_3 \stackrel{\leftrightarrow}D_3 \right) \right) q ~~,
\label{v2}
\end{equation}
\begin{equation}
\cO_{a_0} = 
 \bar q \gamma_2 \gamma_5 q ~~,
\end{equation}
\begin{equation}
\cO_{a_1} = {1\over 4} ~
 \bar q \left( \gamma_4 \gamma_5 \stackrel{\leftrightarrow}D_2
+\gamma_2 \gamma_5 \stackrel{\leftrightarrow}D_4 \right) q ~~,
\end{equation}
\begin{equation}
\cO_{t_0} = 
 \bar q \sigma_{24} \gamma_5 q ~~.
\end{equation}
While the $v_2$ operator Eq.(\ref{v2})
(the $v_{2b}$ operator in the notation of \cite{meinulf5}) is
used in a spin-averaged matrix element, the other operators require a 
polarised baryon state to give a non-zero result.
In these latter cases we choose the spin to point into the 2-direction.
In parton model language we have:
\begin{equation}
v_{2}^q=\langle xq(x)\rangle ~ ~ ~, ~ ~ ~ 
a_0^q=2\langle \Delta q(x)\rangle ~ ~ ~, ~ ~ ~ 
a_1^q=2 \langle x \Delta q(x)\rangle ~ ~ ~, ~ ~ ~ 
t^q_0=2 \langle \delta q(x)\rangle ~~. 
\label{eq2}
\end{equation}
In order to reduce the cut-off effects
from $O(a)$ to $O(a^2)$ also in matrix elements, the improvement of the 
fermionic action has to be accompanied by the improvement of the operator
under study. For the axial vector current  the 
improved renormalised operator has the form 
\begin{equation}
A^{impr}_\mu = Z_{a_0} (1 + b_{a_0} a m) (A_\mu + a c_{a_0} \partial_\mu P)
\end{equation}
with the bare axial vector current 
$A_\mu (x) = \bar{q}(x) \gamma_\mu \gamma_5 q(x)$, the pseudoscalar 
density $P(x) = \bar{q}(x) \gamma_5 q(x)$, and the bare quark mass
$am = 1/(2 \kappa) - 1/(2 \kappa_c)$. The improvement term $\partial_\mu P$
vanishes in forward matrix elements, so we do not need the coefficient
$c_{a_0}$.
For the other operators one can use the  
equation of motion to define the improvement coefficients analogous to 
$c_{a_0}$ such that they are of order $g^2$ and
and thus small.
We therefore decided in the end to neglect the improvement terms altogether.   
For baryons with momentum zero
the ratios $R_{\mathcal O}$ are related to the reduced matrix elements 
(\ref{eq2}) by
\begin{eqnarray}
R_{v_{2}} &=& -\frac{1}{Z_{v_{2}}(1+ b_{v_{2}} a m)} 
        \frac{1}{2 \kappa}
        M v_{2}~~~,
\nonumber \\
R_{a_0} &=& \frac{{\rm i}}{Z_{a_{0}}(1+ b_{a_{0}} a m)} 
        \frac{1}{2 \kappa} \frac{1}{2} a_0 ~~~,
\nonumber \\
R_{a_1} &=& -\frac{{\rm i}}{Z_{a_{1}}(1+ b_{a_{1}} a m)} 
        \frac{1}{2 \kappa} \frac{M}{4}  a_1 ~~~,
\nonumber \\
R_{t_0} &=& \frac{1}{Z_{t_0}(1+ b_{t_0} a m)}
         \frac{1}{2 \kappa}  {1\over 2}t_0 ~~~.
\label{eq3}
\end{eqnarray}
Depending on the flavour of the quark one has to insert for $\kappa$
either $\kappa_d$ or $\kappa_s$.
These relations account for the operator renormalisation and different 
normalisations on the lattice and in the continuum
\cite{meinulf5,meinulf6}.
The renormalisation constants $Z_{v_2},~Z_{a_0},~Z_{a_1},~Z_{t_0}$ 
depend on the renormalisation scale $\mu$ of the continuum theory and the 
lattice cut-off $1/a$ or equivalently $\beta$.
One exception is $Z_{a_0}$, which depends only on $\beta$, because in the 
continuum the 
anomalous dimension is zero due to current conservation.
The scale dependence can be factorized into a factor 
depending on $\mu$ and one depending 
on $\beta$.
For the renormalisation constants we employ a variation of tadpole improved 
perturbation theory  - TRB-PT - as briefly described in 
\cite{r14a}. The renormalisation group invariant form is first found 
(accurate to two loop perturbation theory) and then converted back to
the $\overline{MS}$-scheme at $\mu^2$=4 GeV$^2$, using two-loop 
perturbation theory for
$t_0$ and 
three-loop perturbation theory for $a_1$ and $v_2$. 
We obtain \cite{meinulf7}
\begin{eqnarray}
Z_{v_{2}}(\mu^2=4~{\rm GeV}^2,\beta=6.0) = & 1.11106 
~~~~~~~~& b_{v_{2}} 
= 1.25803 
\nonumber \\
Z_{a_0}(\beta=6.0) = &  0.83232 ~~~~~~~~& b_{a_0} =1.27134 
\nonumber \\
Z_{a_1}(\mu^2=4~{\rm GeV}^2,\beta=6.0) =  &  1.11800 ~~~~~~~~& b_{a_1} =1.24313
\nonumber\\
Z_{t_0}(\mu^2=4~{\rm GeV}^2,\beta=6.0) = & 0.88924  ~~~~~~~~& b_{t_0} =1.24626
 ~~ .\label{eq1}
\end{eqnarray}
$Z_{a_0}$ and $b_{a_0}$ have been determined non-perturbatively 
(on the lattice) in
\cite{losalamos} with the result
\begin{equation}
Z_{a_0}(\beta=6.0) = 0.807 (2)(8) ~ ~ ~ ~ ~ ~ ~  b_{a_0} = 1.28 (3)(4) 
~~~ .
\label{e12}
\end{equation}
This value for $Z_{a_0}$  deviates 
by $\approx 4 \%$ from the perturbative value in Eq.~(\ref{eq1}).
We used for $a_0$ Eq.(\ref{e12}) and otherwise Eq.(\ref{eq1}).
At this point we also want to  address the question with which 
phenomenological distribution functions 
we should compare. Our $Z$ values include all contributions 
of order $\alpha_s$ and some (but not all) contributions of higher
order. 
This suggests that we should compare them with an NLO (or higher) 
DGLAP-fit to the experimental data in the $\overline{MS}$ scheme and for 
the scale 4 GeV$^2$.

\section{RESULTS}

\begin{figure}
\epsfig{file=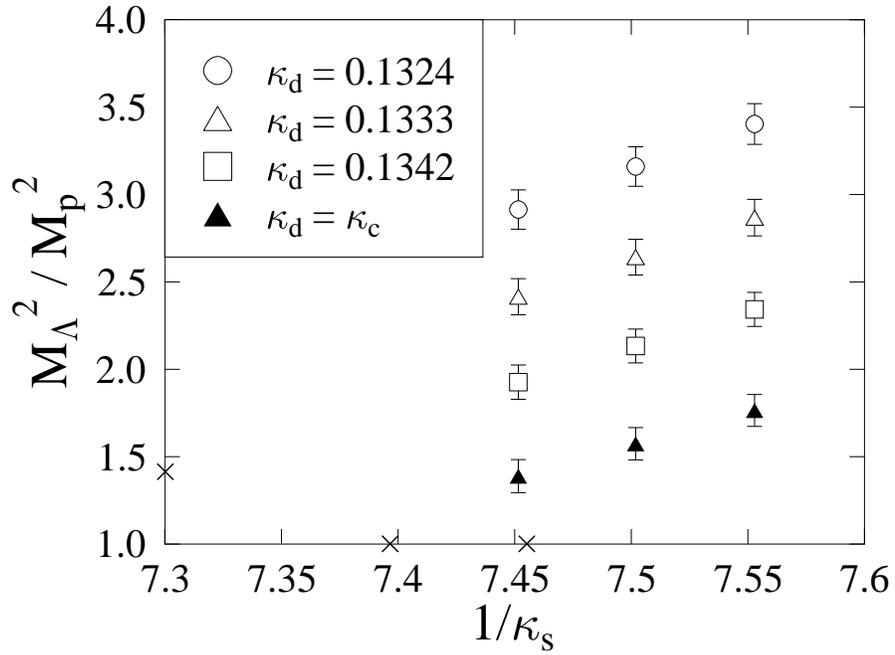,width=14cm} 
\caption{The square of the ratio $M_\Lambda / M_p$ with the proton mass 
$M_p$ taken in the chiral limit versus $1/\kappa_s$. The different
symbols correspond to the different values of $\kappa_d$ including 
the chiral limit. The crosses (left to right) indicate the physical 
value of $M_\Lambda / M_p$, $1/\kappa_c$, and  $1/\kappa_s^*$, respectively.}
\label{fig.mass}
\end{figure}

\begin{figure}
\begin{center}
\epsfig{file=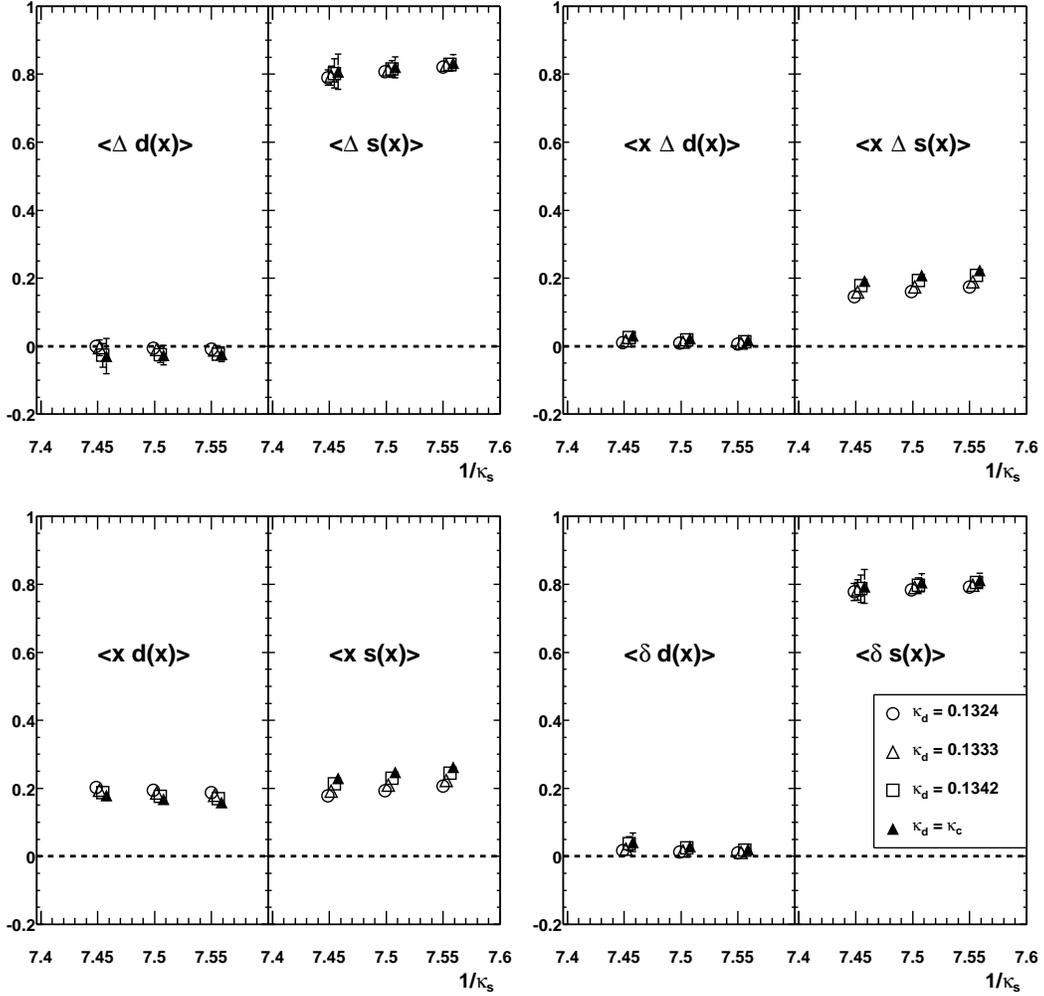,width=14cm} 
\end{center}
\caption{The bare results for
$\langle \Delta q(x) \rangle$, $\langle x\Delta q(x) \rangle$,
$\langle x q(x) \rangle$, and $\langle \delta q(x) \rangle$ in the 
$\Lambda$ versus $1/\kappa_s$.
The results for the different values of $\kappa_d \neq \kappa_c$
have been slightly displaced horizontally.}
\label{fig.delta}
\end{figure}

From the $\Lambda$ masses at our nine combinations of $\kappa_d$, 
$\kappa_s$ we have computed $\Lambda$ masses at $\kappa_d = \kappa_c$
by linear extrapolation of $M_\Lambda^2$ in $1/\kappa_d$.
These 12 masses are plotted in Fig. \ref{fig.mass}.
The filled triangles denote the chiral limit $\kappa_d,$ $\kappa_u$ 
$\rightarrow$ $\kappa_c$  for each $\kappa_s$. Our
value for $\kappa_s^*$, which we fixed by $M_K$,
reproduces quite accurately the ratio 
$M_\Lambda / M_p = 1.19$.
This fits in nicely with the observation that $M_\Sigma / M_p$ and
$M_\Xi / M_p$ are also rather well reproduced by quenched simulations
(see, e.g., \cite{dirk}). The plot shows clearly the breaking of 
the SU(3) flavour symmetry.
In particular, the dependence of the masses on $\kappa_d$ is rather 
pronounced.\\ 
In Fig. \ref{fig.delta} we plot our bare results,
i.e. the results without the factors $Z(1+bam)$ in Eq. (\ref{eq3}),  for 
$\langle \Delta q(x)\rangle$, $\langle x \Delta q(x)\rangle$
$\langle xq(x)\rangle$, and $\langle \delta q(x)\rangle $ 
in the $\Lambda$ versus $1/\kappa_s$
($q=s,d$; the results for $u$ are the same as for  $d$).
We plotted the bare results because these are the numbers we actually used 
for inter- and extrapolation.
In contrast with the case of $M_\Lambda / M_p$, the dependence on $\kappa_d$
is rather weak. The values corresponding to $\kappa_d = \kappa_c$ have been
obtained by extrapolating the bare matrix elements linearly in $1/\kappa_d$
(filled triangles). Finally, by
interpolating the bare matrix elements for $\kappa_d = \kappa_c$
linearly in $1/\kappa_s$ to $1/\kappa_s^*$ we obtain the desired 
$\Lambda$ matrix elements.
The numbers plotted in the figure are given as tables in the appendix.
The physical results for the renormalisation scale $\mu^2=4$ GeV$^2$
are collected in table \ref{tab:t1}. 
\begin{table}
\begin{center}
\begin{tabular}[h] {lll} \hline
{} & \multicolumn{1}{c}{$u,d$} & 
    \multicolumn{1}{c}{$s$}\\ \hline 
$\langle x q(x)\rangle$ &
   \multicolumn{1}{c}{$0.20(1)$}     & \multicolumn{1}{c}{$0.27(1)$}   \\
$\langle \Delta q(x)\rangle$ &
   \multicolumn{1}{c}{-$0.02(4)$}   & \multicolumn{1}{c}{$0.68(4)$}    \\
$\langle x \Delta q(x)\rangle$&
   \multicolumn{1}{c}{$0.033(8)$}    & \multicolumn{1}{c}{$0.23(2)$}      \\
$\langle \delta q(x)\rangle$&
   \multicolumn{1}{c}{$0.03(2)$}     & \multicolumn{1}{c}{$0.74(4)$}      \\
   \hline
\end{tabular}
\caption{\label{tab:t1} Our results for different moments of polarised and 
unpolarised quark distribution functions in the $\overline{MS}$ scheme
at $\mu^2=4$ GeV$^2$.} 
\end{center}
\end{table}
Thus we find that the $d$ and $u$ quarks carry a somewhat smaller 
momentum fraction than the $s$ quarks, 20\%  each as compared to
27\% which is quite intuitive in view of the larger 
$s$ mass. 
We also find that the tensor charge $\delta q$
is close to $\Delta q$ for the $s$ quark,
which again fits to our understanding of transversity,
as both should be identical in the non-relativistic limit.
Note that a positive value of 
$\langle x\Delta d(x) \rangle$ and a 
smaller or even negative value for 
$\langle \Delta d(x) \rangle$ imply a sign change of $\Delta d(x)$ 
as function of $x$. 
This would agree with model predictions \cite{r1}.\\
These numbers are subject to systematic uncertainties which we will now 
discuss. 
For the nucleon similar quenched lattice calculations always
gave  quark momentum fractions which were too large by typically
about 20\% and this disagreement was usually attributed to
quenching and the neglect of quark-line disconnected 
contributions. 
Recently it has been pointed out that the 
extrapolation
to the chiral limit could be of similar or even greater importance
\cite {hemmert}. If this is true we expect that 
the corrections from a nontrivial chiral extrapolation can affect 
substantially the values for $\Delta q$. 
Another source of uncertainty is the lacking continuum extrapolation as the 
current calculations have been performed at one lattice spacing only.\\
Let us now discuss our results for $\Delta q$ in more detail on the basis 
of table \ref{tab:t2}.
\begin{table}
\begin{center}
\renewcommand{\arraystretch}{1.4}
  \begin{tabular}[h]{lll} \hline 
     {}  & \multicolumn{1}{c}{$\Delta u_\Lambda = \Delta d_\Lambda$} & 
    \multicolumn{1}{c}{$\Delta s_\Lambda$}\\ \hline 
    quark model & 
   \multicolumn{1}{c}{$0$}         & \multicolumn{1}{c}{$1$}      \\
    exp. + SU(3)$_{\mathrm F}$ & 
          $-0.17(3)$                    &          $0.63(3)$           \\
    MC + SU(3)$_{\mathrm F}$ &
            $-0.016(9)$               &         $0.65(2)$               \\
    this work   &
                 $-0.02(4)$       &        $0.68(4)$                 \\
   \hline
  \end{tabular}
\caption{\label{tab:t2} Comparison of our results for the longitudinal
quark polarisation in the $\Lambda$ with the naive quark model prediction, 
the flavour SU(3) rotated experimental values for the proton and
flavour rotated lattice results for the proton.}
\end{center}
\end{table}
In this table we give
in the third line  the $\Lambda$ matrix elements as they follow from
our Monte Carlo results for the {\em proton} matrix elements by the use of 
SU(3)$_{\mathrm F}$ (see Eq.(3)). They agree quite well with the matrix 
elements computed directly (fourth line). This implies that 
the flavour symmetry breaking effects in the
matrix elements are rather small,
which can also be concluded from Fig.~\ref{fig.delta}.
As the mass difference between the light quarks and the strange 
quark was consistently taken into account this provides a 
strong argument that the $\Lambda$ and proton spin structures are, 
in good 
approximation, simply related by an SU(3) transformation.
Consequently, the values given in the second line, which were  
computed from the proton spin structure under the assumption of flavour
SU(3)(see, e.g., \cite{ashery}), should be quite reliable.\\
Because we performed a quenched calculation it might be more 
consistent to compare our results to 
the prediction for the 
valence quark contribution, e.g. by Ashery and 
Lipkin~\cite{ashery}. They obtained  
$ \Delta u_\Lambda = \Delta d_\Lambda = -0.07(4)$, 
$ \Delta s_\Lambda = 0.73(4)$. 
Notice that
all  results differ markedly from 
the predictions of the (naive) quark model shown in the first line.

\section{SUMMARY}

We have studied various aspects of the $\Lambda$ (spin) structure taking
the mass difference between the quarks into account. We found that
our results agree nicely with general expectations: The momentum
fraction carried by the $s$ quark is larger
than for light quarks, the values for 
the tensor charge $\delta s$ and $\Delta s$ are similar. 
Our main result is, however, that 
SU(3) flavour symmetry appears to be much
less violated in the matrix elements than in
the baryon masses, which is in agreement with the empirical
observation that the hyperon semileptonic
decays can be parametrised rather well assuming SU(3) flavour symmetry.
Presently  the use of 
$\Lambda$s as a probe of the nucleon spin structure in e.g.
deep-inelastic lepton-nucleon scattering 
still has to rely on models, but the properties of the
latter should definitely 
be checked by a comparison  with our lattice results.\\
We expect that our results have a similar uncertainty as those for
the (spin) structure of the 
proton and that the origin of this uncertainty is the same. 
When configurations with lighter dynamical quarks become available 
and our understanding of chiral extrapolation improves, high precision  
results for the $\Lambda$ should become reachable along the lines discussed 
in this contribution.

\section*{ACKNOWLEDGEMENTS}
This work is supported in part by the DFG 
(Schwerpunkt ``Elektromagnetische Sonden'')
by BMBF and by the European Community's Human Potential Program
under contract HPRN-CT-2000-00145, Hadrons/Lattice QCD.
The numerical calculations were performed on the Quadrics 
computers at DESY Zeuthen. We wish to thank the operating staff 
for their support.

\section*{APPENDIX}
The following tables contain the numerical values for our 
lattice results.

\begin{table}
\begin{center}
\begin{tabular}{p{0.12\textwidth}p{0.12\textwidth}||p{0.12\textwidth}p{0.12
\textwidth}
||p{0.12\textwidth}p{0.12\textwidth}}
$\kappa_d$ & $\kappa_s$ & $a_0^d$  &  $a_0^s$ & $a_1^d$  &  $a_1^s$\\
\hline
\hline
$0.1324$&$0.1324$&$-0.02(1)$&$1.64(2)$&$0.014(4)$&$0.350(7)$\\
$0.1333$&$0.1324$&$-0.02(2)$&$1.65(3)$&$0.019(5)$&$0.378(9)$\\
$0.1342$&$0.1324$&$-0.05(4)$&$1.66(4)$&$0.027(7)$&$0.416(15)$\\
$0.1324$&$0.1333$&$-0.01(2)$&$1.61(3)$&$0.017(4)$&$0.321(8)$\\
$0.1333$&$0.1333$&$-0.02(3)$&$1.63(3)$&$0.024(6)$&$0.350(10)$\\
$0.1342$&$0.1333$&$-0.05(5)$&$1.63(5)$&$0.037(9)$&$0.387(18)$\\
$0.1324$&$0.1342$&$-0.00(3)$&$1.58(4)$&$0.022(6)$&$0.291(9)$\\
$0.1333$&$0.1342$&$-0.01(5)$&$1.59(6)$&$0.032(7)$&$0.318(13)$\\
$0.1342$&$0.1342$&$-0.05(7)$&$1.60(9)$&$0.051(12)$&$0.357(27)$\\
\hline
$\kappa_\mathrm{c}$&$0.1324$&$-0.05(4)$&$1.67(5)$&$0.03(1)$&$0.45(2)$\\
$\kappa_\mathrm{c}$&$0.1333$&$-0.05(6)$&$1.64(6)$&$0.05(1)$&$0.42(2)$\\
$\kappa_\mathrm{c}$&$0.1342$&$-0.06(10)$&$1.61(10)$&$0.06(2)$&$0.38(3)$\\
\hline
$\kappa_\mathrm{c}$&$\kappa_s^*$&$-0.06(9)$&$1.62(9)$&$0.06(2)$&$0.39(3)$
\end{tabular}
\caption{\label{tab:a0} Measured bare values for $a_0$ and $a_1$.}
\vspace{1cm}
\begin{tabular}{p{0.12\textwidth}p{0.12\textwidth}||p{0.12\textwidth}p{0.12
\textwidth}
||p{0.12\textwidth}p{0.12\textwidth}}
$\kappa_d$ & $\kappa_s$ & $v_2^d$  &  $v_2^s$ & $t_{0}^d$  &  $t_{0}^s$\\
\hline
\hline
$0.1324$&$0.1324$&$0.188(2)$&$0.207(3)$& $ 0.02(1)$ & $ 1.58(2) $\\
$0.1333$&$0.1324$&$0.179(3)$&$0.223(4)$& $ 0.02(1)$ & $ 1.60(3) $\\
$0.1342$&$0.1324$&$0.169(5)$&$0.244(5)$& $ 0.04(2)$ & $ 1.61(3) $\\
$0.1324$&$0.1333$&$0.194(3)$&$0.193(4)$& $ 0.03(1)$ & $ 1.57(3) $\\
$0.1333$&$0.1333$&$0.186(4)$&$0.209(4)$& $ 0.03(1)$ & $ 1.58(3) $\\
$0.1342$&$0.1333$&$0.177(7)$&$0.230(6)$& $ 0.05(3)$ & $ 1.59(4) $\\
$0.1324$&$0.1342$&$0.202(3)$&$0.177(5)$& $ 0.03(1)$ & $ 1.55(5) $\\
$0.1333$&$0.1342$&$0.195(4)$&$0.192(6)$& $ 0.04(2)$ & $ 1.57(6) $\\
$0.1342$&$0.1342$&$0.188(9)$&$0.213(9)$& $ 0.08(4)$ & $ 1.57(8) $\\
\hline
$\kappa_\mathrm{c}$&$0.1324$&$0.159(7) $&$0.262(6)$  & $0.04(3)$ &
$1.62(4)  $\\
$\kappa_\mathrm{c}$&$0.1333$&$0.168(8) $&$0.248(7)$  & $0.06(4)$ &
$1.61(5)  $\\
$\kappa_\mathrm{c}$&$0.1342$&$0.180(11)$&$0.229(10)$   & $0.08(5)$ &
$1.59(10) $ \\
\hline
$\kappa_\mathrm{c}$&$\kappa_s^*$&$0.178(10)$&$0.233(10)$ & $ 0.08(5) $ &
$1.59(8)  $
\end{tabular}
\end{center}
\caption{\label{tab:v2} Measured bare values for $v_2$ and $t_{0}$.}
\end{table}

\newpage

\end{document}